\pdfoutput=1
\documentclass{article}

\PassOptionsToPackage{numbers, compress}{natbib}

\usepackage[final]{neurips_2022_ml4ps}

\usepackage[utf8]{inputenc} 
\usepackage[T1]{fontenc}    
\usepackage{hyperref}       
\usepackage{url}            
\usepackage{booktabs}       
\usepackage{amsfonts}       
\usepackage{nicefrac}       
\usepackage{microtype}      
\usepackage{xcolor}         
\usepackage{amsmath}
\usepackage{graphicx}

\title{Ad-hoc Pulse Shape Simulation using Cyclic Positional U-Net}

\author{%
  Aobo Li\thanks{Corresponding Author} \\
  University of North Carolina at Chapel Hill,\\
  Triangle Universities Nuclear Laboratory\\
  \texttt{liaobo77@ad.unc.edu} \\
  \And
  Julieta Gruszko \\
  University of North Carolina at Chapel Hill,\\
  Triangle Universities Nuclear Laboratory\\
  \texttt{jgruszko@unc.edu} \\
  \AND
  Brady Bos \\
  University of North Carolina at Chapel Hill,\\
  Triangle Universities Nuclear Laboratory\\
  \texttt{bradybos@live.unc.edu} \\
  \And
  Thomas Caldwell \\
  University of North Carolina at Chapel Hill,\\
  Triangle Universities Nuclear Laboratory\\
  \texttt{tcald@unc.edu} \\
  \And
  Esteban Le\'on \\
  University of North Carolina at Chapel Hill,\\
  Triangle Universities Nuclear Laboratory\\
  \texttt{esleon97@live.unc.edu} \\
     \And
  John Wilkerson \\
  University of North Carolina at Chapel Hill,\\
  Triangle Universities Nuclear Laboratory\\
  \texttt{jfw@unc.edu} \\
}

\begin{document}

\maketitle

\begin{abstract}
High-Purity Germanium~(HPGe) detectors have been a key technology for rare-event searches, such as neutrinoless double-beta decay and dark matter searches, for many decades. Pulse shape simulation is pivotal to improving the physics reach of these experiments. In this work, we propose a Cyclic Positional U-Net~(CPU-Net)\footnote{\url{https://github.com/aobol/CPU-Net.git}} to achieve ad-hoc pulse shape simulations with high precision and low latency. Taking the transfer learning approach, CPU-Net translates simulated pulses to detector pulses such that they are indistinguishable. We demonstrate CPU-Net's performance on data taken from a local HPGe detector.
\end{abstract}

\section{Introduction}\label{sec:intro}
Ge detectors have been at the heart of low-background experimental searches for neutrinoless double-beta decay and dark matter~\cite{IGEX,COGENT,MJD,legend_pcdr,SuperCDMS,GERDA,CDMS}. When a particle interacts in the active volume of a depleted Ge detector, a large number of electron/hole pairs are created and drift to the electrodes. In point-contact-style detectors, the signal is read out at a small point-like ground electrode, with the resulting electric field leading to sharply-rising detector pulses. This readout electrode is instrumented with an amplification and digitization chain. In many experiments, the first stage of amplification is provided by a charge-sensitive resistive feedback preamplifier, leading to a slowly-falling tail following the pulse \cite{MJD_electronics}. 

The shape of the detector pulses is affected by many parameters, including particle energy, interaction position, charge cloud size, and so on. Pulse shape simulation aims at generating pulses based on these parameters that are indistinguishable from actual detector pulses. This allows us to produce ground-truth datasets where a bijective mapping is established between each simulated pulse and its corresponding incident particle(s). Suppose we apply an existing pulse shape discrimination cut~\cite{AvsE} to this dataset. In that case, It will be immediately apparent which event topologies lead to the signal sacrifice or background contamination of this cut. This information will be pivotal for the evaluation and further improvement of cut efficiency. Pulse shape simulation is also critical for designing new cuts and new HPGe detector geometries, as well as training machine learning models.

Traditional pulse shape simulation, which can be conducted with publicly-available tools such as the \texttt{fieldgen} and \texttt{siggen}\footnote{\url{https://github.com/radforddc/icpc_siggen}} or \texttt{SSD}\footnote{\url{https://github.com/JuliaPhysics/SolidStateDetectors.jl}} software packages, starts from an electric field simulation that takes into account the geometry and impurity concentration of each detector. Each energy deposition, taken from a particle interaction simulation tool such as Geant4 \cite{geant4one,geant4two}, is used to simulate the pulse induced by charge drift and collection from the deposition site. 
The pulses are scaled according to the energies deposited at each site, and summed together to create the total event waveform. A series of corrections based on physical first principles must then be added to reproduce the detector response. Those first-principle corrections---including charge trapping~\cite{charge_trapping}, surface effects from the high voltage electrode and insulating surfaces ~\cite{dl}~\cite{TUBE_paper}, and the electronic response of the readout chain~\cite{Ben_Thesis,Sam_Thesis}, are hard to model \textit{a priori}. They vary detector-by-detector, and can also vary as a result of the operating conditions of the experiment, which may change over time. This makes high-fidelity pulse shape simulation extremely challenging; thus, all current-generation experiments have avoided it by directly calculating reconstruction parameters from Monte Carlo particle-interaction simulations with heuristic methods for most of their simulation needs.

Transfer learning lifts the ``curse of the first principle'' by directly learning the ad-hoc translations between each simulated pulse and its corresponding detector pulse. In this work, we present a novel neural network model called Cyclic Positional U-Net~(CPU-Net) for ad-hoc pulse shape simulation. CPU-Net translates simulated pulses to detector pulses in arbitrarily collected datasets so that they are indistinguishable according to the shape and reconstruction parameter distributions. The data-driven nature of CPU-Net allows fast training and straightforward generalization to multiple detectors and changing operating conditions without detector-wise model tuning. We also confirm that CPU-Net has learned the proper detector physics to conduct the translation without explicit programming.

\section{Cyclic Positional U-Net}~\label{sec:cpu_net}
Simulation tasks can be re-formulated under the transfer learning framework. Suppose there is a source domain:
\begin{equation}
    \mathcal{D}_{Source}=\{\mathcal{Y},P(Y)\}\qquad Y=\{\mathrm{E},I_{\mathrm{max}},c_{\mathrm{tail}}...\}\in \mathcal{Y}
    \label{eqn:source_domain}
\end{equation}
containing a feature space $\mathcal{Y}$ following a certain probability distribution $P(Y)$. $Y$ contains the reconstruction parameters---energy, maximal current amplitude, tail slope and so on---of the pulse to be simulated. The set of simulated pulses $\mathcal{X}$ can be described as the source task:
\begin{equation}
    \mathcal{T}_{Source}=\{\mathcal{X},P(X|Y)\} \qquad X\in \mathcal{X}
    \label{eqn:source_task}
\end{equation}
On the other hand, the target domain $\mathcal{D}_{Target}=\{\mathcal{Y}',P(Y')\}$ contains the reconstruction parameters of detector pulses, therefore $\mathcal{T}_{Target}=\{\mathcal{X}',P(X'|Y')\}$ represents the detector pulses themselves. Pulse shape simulation aims to learn $P(X'|Y')$ and apply it to an arbitrary $Y$ in $\mathcal{D}_{Source}$ so that the generated $\mathcal{X}$ is indistinguishable from $\mathcal{X}'$. Traditional method attempts to model $P(X'|Y')$ by introducing nuisance parameters into $P(X|Y)$ and fitting $\mathcal{T}_{Target}$ to obtain their values. The nuisance parameter design requires complicated modeling and characterization data-taking, along with computationally expensive fitting procedures in high-dimensional, highly degenerate parameter space~\cite{Ben_Thesis,Sam_Thesis}. In this work, we avoid the direct modification of  $P(X|Y)$ by introducing the Ad-hoc Translation Network~(ATN):
\begin{equation}
\Lambda = \{\hat{\mathcal{X}}, P(\hat{X}\mid X)\}\qquad \hat{X}\in \hat{\mathcal{X}}
\label{eqn:atn}
\end{equation}
ATN accepts an input pulse $X$ and translates it to an output pulse $\hat{X}$. The transformation to be applied is learned from the data, and the collection of transformed output $\hat{\mathcal{X}}$ should be indistinguishable from $\mathcal{X}'$ after training. Therefore, by combining the ATN and $\mathcal{T}_{Source}$, we can replicate $\mathcal{T}_{Target}$:
\begin{equation}
    \mathcal{T}_{Target}=\Lambda \mathcal{T}_{Source} ,
    \label{eqn:atn_task}
\end{equation}
which allows us to precisely reproduce $\mathcal{T}_{Target}$ without modifying  $P(X|Y)$.

We choose a 1D U-Net~\cite{UNet} as the baseline model when designing the ATN. U-Net contains $n$ convolutional layers to encode each pulse into a feature vector, then uses $n$ upsample layers to decode the feature vector back to an output with the same length. In addition, $n$ contracting paths are established between each pair of convolutional layers and upsample layer at the same level. This network structure allows information at different levels to flow to the decoding part, providing maximal reconstruction efficiency. 

\begin{figure}[htb!]
    \includegraphics[width=0.4\linewidth,trim={0pc 29pc 5pc 4pc},clip]{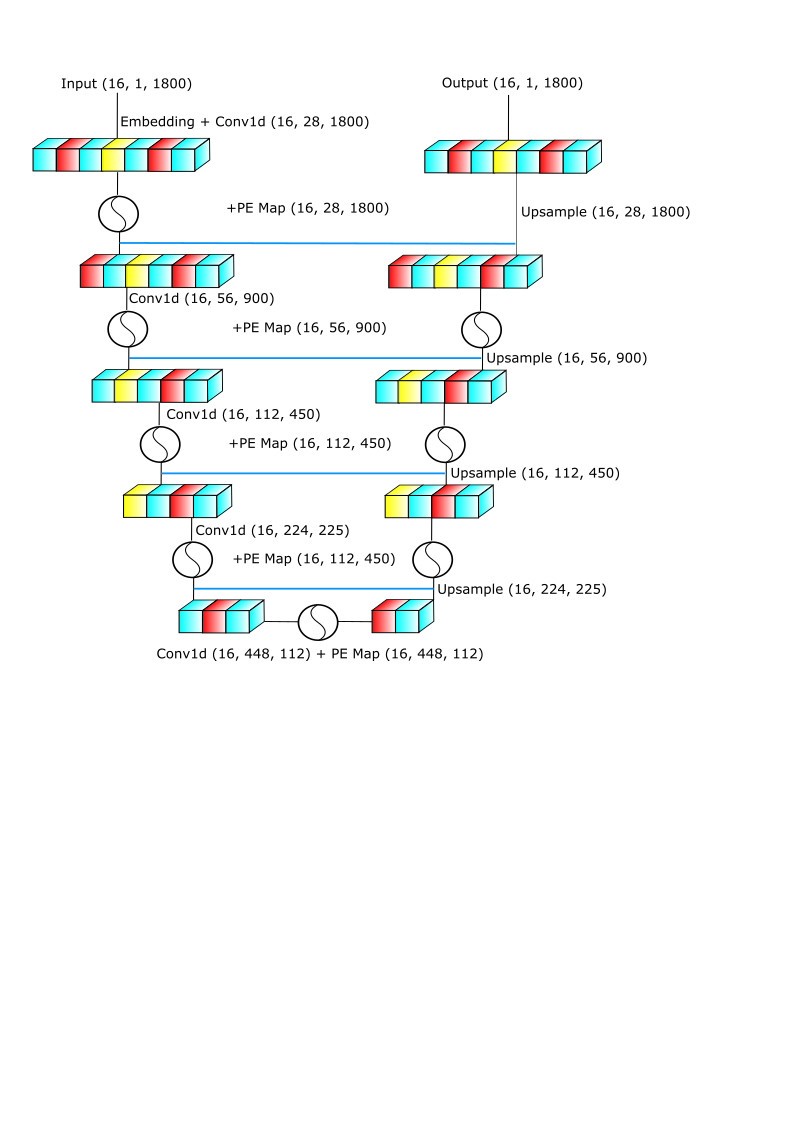}
    \hspace{0.05\linewidth}
    \includegraphics[width=0.45\linewidth,trim={0pc 0pc 0pc 0pc},clip]{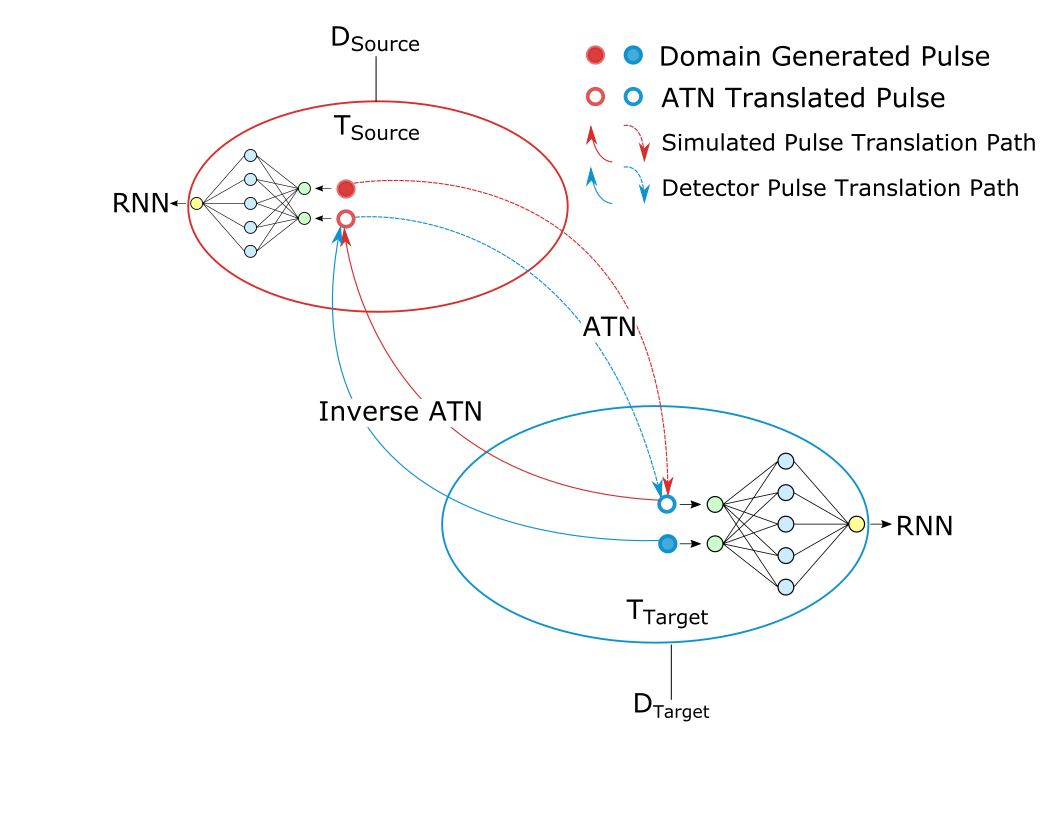}
    \label{fig:cpunet}
    \caption{(Left) Schematic diagram of Positional U-Net~(PU-Net). The blue line represents the contracting path. Positional encoding maps at the same level are of the same shape. (Right) Cycle-consistent adversarial training with PU-Net as the Ad-hoc Translation Network~(ATN) and Inverse Ad-hoc Translation Network~(IATN). The red oval region represents simulated pulses in $\mathcal{T}_{Source}$, and the blue oval region represents detector pulses in $\mathcal{T}_{Target}$. }
   \label{fig:network_schematic}
\end{figure}

During network design, we observed that conventional U-Net does not reproduce the tail of the waveform (the cyan region in Figure~\ref{fig:result} top panel) due to the lack of positional information in intermediate layer outputs. Therefore, we proposed a Positional U-Net~(PU-Net) model with layer-wise positional encoding maps $\mathcal{M}_{\mathrm{position}}$ inspired by the Transformer model~\cite{Transformer}. $\mathcal{M}_{\mathrm{position}}$ contains sine and cosine functions with different frequencies. Since each U-Net layer outputs a tensor with a different shape, $\mathcal{M}_{\mathrm{position}}$ must be generated separately for each layer and added to the layer output, as shown in the left panel of Figure~\ref{fig:network_schematic}.

Ideally, the ATN should be trained with paired pulses ($X, X'$) where $Y=Y'$ for each pair. In reality, it is impossible to collect such a paired dataset: we can simulate $\mathcal{T}_{Source}$ from an arbitrary $\mathcal{D}_{Source}$, but it is impossible to collect $\mathcal{T}_{Target}$ so that $\mathcal{D}_{Source} = \mathcal{D}_{Target}$ without knowing the exact form of $P(X'|Y')$. Therefore, the training has to be conducted on $\mathcal{T}_{Source}$ and $\mathcal{T}_{Target}$ given that $\mathcal{D}_{Source} \neq \mathcal{D}_{Target}$. The CycleGAN framework~\cite{CycleGAN} allows us to train ATN on such an unpaired dataset. We first construct two networks ---an ATN~ $\Lambda$ and an Inverse ATN $\Bar{\Lambda}$, both with PU-Net structure. We then construct two discriminator network $\delta_{S}$ and $\delta_{T}$ for the source and target pulses, respectively. We choose attention-coupled recurrent neural network~(RNN)~\cite{attention} as the discriminator model because position is intrinsically enforced by RNN. When training the network, a simulated pulse $X$ is first fed to $\Lambda$ to produce a translated pulse $\Lambda(X)$, then an adversarial training with $\delta_{T}$ is performed: $\delta_{T}$ attempts to distinguish $\Lambda(X)$ from $\mathcal{X}'$ while $\Lambda$ attempts to ``fool'' $\delta_{T}$. Then $\Lambda(X)$ is fed to $\Bar{\Lambda}$ to translate back to $\mathcal{X}$. A cycle-consistent loss ensures the circular translation path preserves the original pulse shape:
\begin{equation}
    L_{\mathrm{Cycle}} = \lvert X - \Bar{\Lambda}(\Lambda(X))\lvert
\end{equation}
The $\mathcal{X}\xrightarrow[]{}\Lambda(\mathcal{X})\xrightarrow[]{}\Bar{\Lambda}(\Lambda(\mathcal{X}))$ translation path is denoted by the red arrows in the right-side panel of Figure~\ref{fig:network_schematic}. Alternatively, the $\mathcal{X}'\xrightarrow[]{}\Bar{\Lambda}(\mathcal{X}')\xrightarrow[]{}\Lambda(\Bar{\Lambda}(\mathcal{X}'))$ is denoted by blue arrows in the same figure. Each path will produce both an adversarial loss $L_{\mathrm{Advesarial}}$ and a cycle-consistent loss $L_{\mathrm{Cycle}}$. The four losses from the two paths are minimized simultaneously during training.

Combining all these, we obtain the Cyclic Positional U-Net~(CPU-Net) for Ad-hoc pulse shape simulation. The trained CPU-Net produces both an ATN and an IATN, translating pulses between the source and target domain. The ATN is the primary interest of this work, but the IATN can also be adopted to boost the physical analysis.

\section{Training Results}\label{sec:result}
\begin{figure}[htb!]
    \includegraphics[width=0.9\linewidth,trim={15pc 2pc 15pc 5pc},clip]{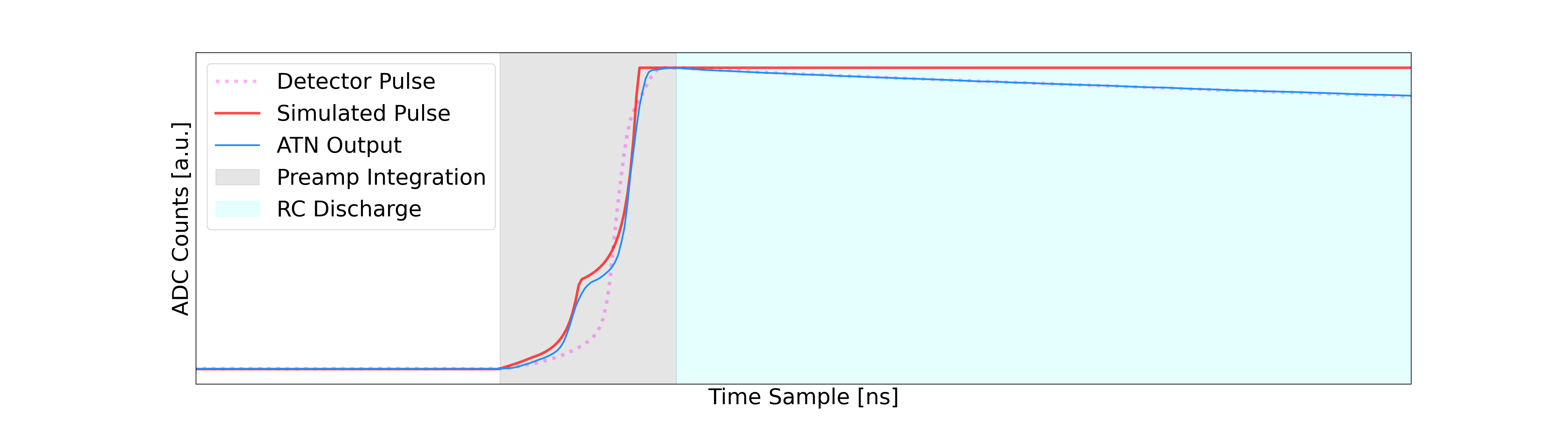}
    \includegraphics[width=0.46\linewidth,trim={0pc 1.5pc 3pc 4pc},clip]{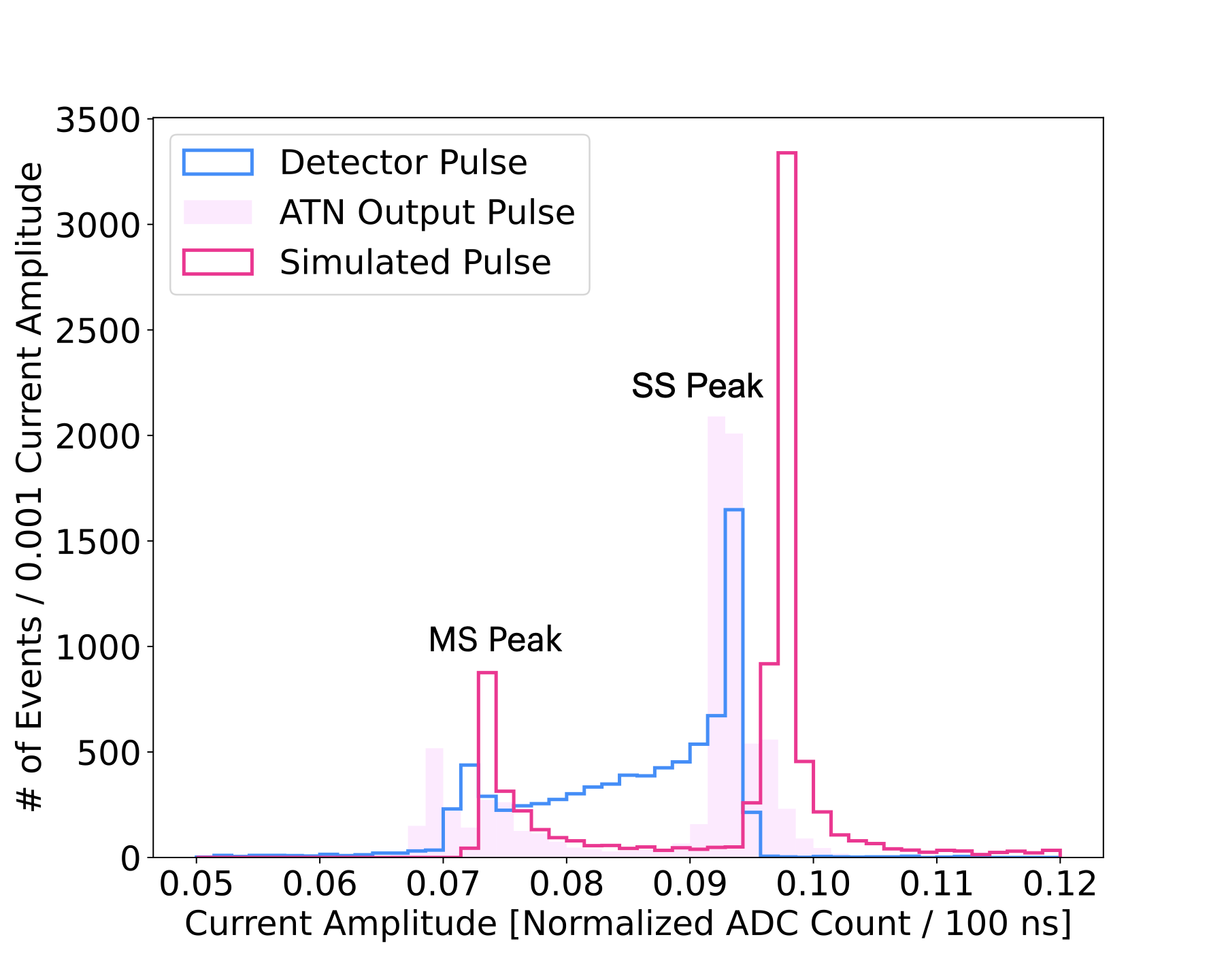}
    \includegraphics[width=0.46\linewidth,trim={0pc 1.5pc 2pc 4pc},clip]{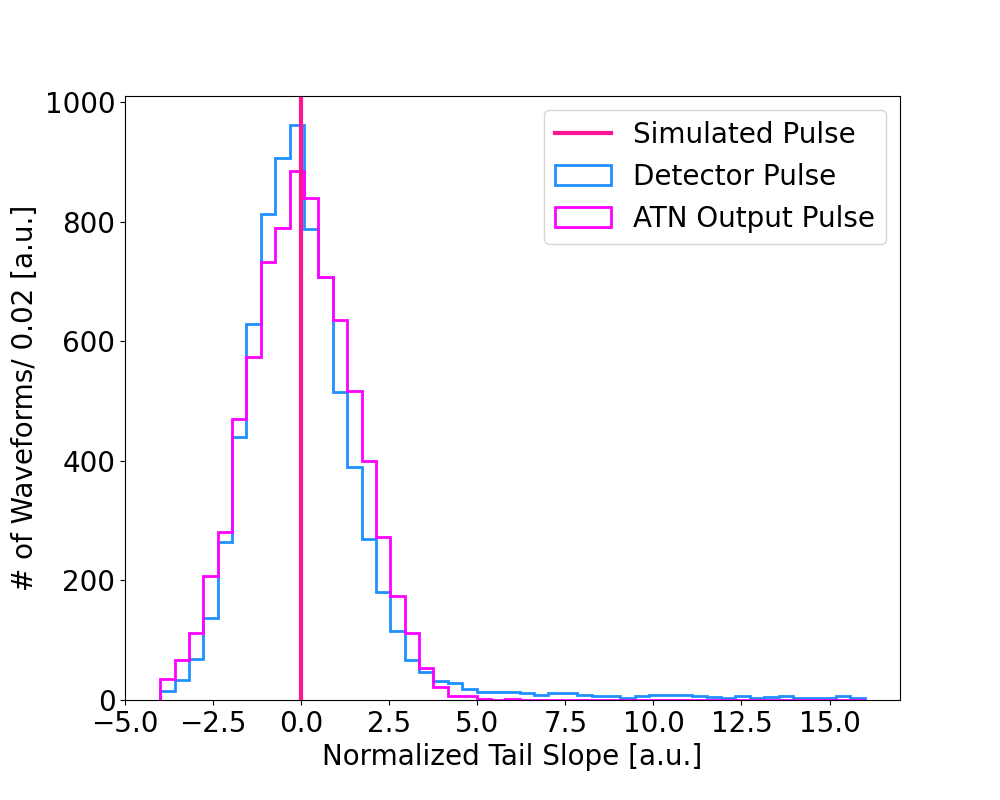}
    \caption{(Top) An example ATN output (blue) generated by the input simulated pulse (red). A detector pulse (dotted magenta,randomly drawn from data) is also illustrated as a reference. The grey region depicts the area where the preamplifier integration effect is most visible. The blue region shows the impact of the RC circuit discharge effect. (Lower Left) A comparison of the maximum current amplitude ($I_{max}$) distribution in different datasets. The two peaks correspond to SS and MS waveforms. (Lower Right) The distribution of the normalized tail slope ($c_{tail}$) in different datasets.}
   \label{fig:result}
\end{figure}

Two major waveform categories that low-background HPGe experiments seek to distinguish are single-site~(SS) and multi-site~(MS) events. When a physical event deposits all its energy in a single location in a detector, a SS waveform with a single sharply-rising step is produced. Conversely, a MS waveform with multiple steps is produced if energy is deposited at multiple locations within the crystal. SS and MS waveforms of the same total energy can be efficiently distinguished by the maximal current amplitude reconstruction parameter $I_{max}$. For a given event energy, SS waveforms tend to have faster $I_{max}$ than MS waveforms, as shown in the lower left panel of Figure~\ref{fig:result}.

To train and validate the performance of the CPU-Net on both SS and MS waveforms, we collected simulated pulses and detector pulses to form $\mathcal{T}_{Source}$ and $\mathcal{T}_{Target}$, respectively. $\mathcal{T}_{Source}$ is generated by \texttt{siggen}. 65,536 SS waveforms are uniformly simulated at different positions of the detector's active volume. Part of the dataset is converted to MS waveforms by randomly stacking two SS waveforms together. The detector pulses are collected using a $^{228}$Th radioactive source, which is attached to the copper cryostat system containing a Broad Energy Ge~(BEGe) detector. We then use an energy cut between 2094-2104\,keV to select Single Escape Peak~(SEP) events as $\mathcal{T}_{Target}$. Over $90\%$ of SEP events are multi-site; taking into account the additional background events in this energy window, $\mathcal{T}_{Target}$ is composed of $\sim 79\%$ MS waveforms. The collected datasets are then split into training and validation subset with a 7:3 ratio.

The training and validation of CPU-Net are conducted in PyTorch~\cite{pytorch}. The result is illustrated in Figure~\ref{fig:result}. As shown in the top panel, the ATN translates the simulated pulse to the ATN output by smoothing the sharp turning edge in the grey region. This is a consequence of the non-zero integration time in detector pulses, which is set to 0 in \texttt{siggen}. The RC discharge effect of the electronic readout system is also set to 0 in \texttt{siggen}, leading to a tail slope of 0 in the simulated waveforms. The ATN learns to translate the flat tail in the cyan region into an exponential decay, and the strength of the decay can be measured by the tail slope reconstruction parameter $c_{tail}$. These translations can be examined on distribution level by plotting the histogram with respect to key reconstruction parameters. We then provide a quantitive evaluation by calculating the Histogram Intersection over Union~(HIoU) between each dataset and the detector pulses. As shown in the lower right panel, the ATN successfully reproduce the $c_{tail}$ distribution of detector pulses, increasing HIoU from 0 on simulated pulses to $83.2\substack{+5.4 \\ -21.7}\%$ on ATN outputs. As shown in the lower left panel, the ATN also reproduced the $I_{max}$ by increasing HIoU from $12.4\%$ to $36.8\substack{+4.2 \\ -8.5}\%$. The uncertainty calculation is discussed in Appendix~\ref{App:uncertainty}.

\section{Conclusion}\label{sec:conclusion}
Pulse shape simulation is pivotal to understanding the detector physics within germanium crystals. Unfortunately, traditional pulse shape simulation is inefficient due to the complex effects inherent to the detector technology that are difficult to model. In this work, we presented the Cyclic Positional U-Net to achieve ad-hoc pulse shape simulation without modeling any of those effects. This model trains on arbitrarily collected and unpaired datasets of simulated and detector pulses. CPU-Net successfully translates simulated pulses to outputs indistinguishable from actual detector pulses by applying corrections according to proper detector physics. Furthermore, we showed that the CPU-Net correctly reproduces the distribution of two critical pulse shape reconstruction parameters. 

\section{Broader Impact}\label{sec:broader_impact}
This section describes the broader impact of this work, including its limitations, future work and application, as well as social impact.

\textbf{Limitations} The major limitation of this work is related to its ability to translate waveforms. Although we have confirmed that CPU-Net learned the proper detector physics, the ATN still occasionally produces local, non-physical responses such as a small but sharp ``spike'' on the output pulses. This could be a consequence of the numerical effects~(such as activation function or positional encoding addition) within the U-Net structure. On the distribution level, the ATN did not capture the minority event population in the long-tailed dataset. As shown in Figure~\ref{fig:result} lower right, the distribution of high $c_{tail}$ events~($c_{tail}>5.0$) are not properly reproduced. The possible cause would be: a) CPU-Net failed to learn from this tail event population b) these events are missing in \texttt{siggen} simulation. Further investigation is needed to find the exact reason. The last limitation is that CPU-Net occasionally fails to train and in these cases, the HIoU is much worse than the result in Figure~\ref{fig:uncert}. As far as we are aware, the training of adversarial models is difficult, since there is no guarantee that the adversarial opponents will eventually reach the Nash equilibrium. Therefore, occasional failure of adversarial training is expected.

\textbf{Future Work and Application} CPU-Net is designed for HPGe detectors. Therefore we plan to apply it to next-generation HPGe detector experiments such as LEGEND~\cite{legend_pcdr}. Furthermore, we plan to investigate and attempt to fix the above-mentioned limitations. We also plan to train CPU-Net on other datasets, such as the characterization dataset of surface events, where the position of events in data is known to better accuracy.

\textbf{Social Impact} CPU-Net is a natural tool to be applied to any semiconductor detector-based experiment. Furthermore, since CPU-Net trains on unpaired time series data, any experiments producing data in waveform format~(e.g. bolometer experiments~\cite{CUORE}) can adopt this model to boost the simulation accuracy. The main negative impact is the energy consumption to the environment of taking HPGe detector data and training the network on CPUs.

\section{Acknowledgement}\label{sec:ack}
This material is based upon work supported by the U.S. Department of Energy, Office of Science, Office of Nuclear Physics, under Award Number DE-SC0022339. This work is done in support of the \textsc{Majorana Demonstrator} and the LEGEND experiment and we thank our collaborators for their input. This work was performed in part at the Aspen Center for Physics, which is supported by National Science Foundation grant PHY-1607611. We also thank Dr. Michelle Kuchera for the inspiring conversation.
\bibliographystyle{plain}
\bibliography{CPUNet}

\newpage
\section{NeurIPS Checklist}
For all authors...
\begin{itemize}
    \item Do the main claims made in the abstract and introduction accurately reflect the paper's contributions and scope? [Yes] {\color{blue} the paper's contrubution and scope is discussed in the Abstract, Section~\ref{sec:intro} and Section~\ref{sec:conclusion}.}
    \item Have you read the ethics review guidelines and ensured that your paper conforms to them? [Yes] \textit{We have read the ethics review guidelines and made sure that our paper certainly conforms. All data we used in this work is taken from a local HPGe detector, we carefully followed the environmental health and safety requirement during data taking.}
    \item Did you discuss any potential negative societal impacts of your work? [Yes] {\color{blue}Social Impact is discussed in Section~\ref{sec:broader_impact}}
    \item Did you describe the limitations of your work? [Yes] {\color{blue}Limitation is discussed in Section~\ref{sec:broader_impact}}
\end{itemize}
If you are including theoretical results... [No]  {\color{blue}This paper does not contain any theoretical results}
\begin{itemize}
    \item Did you state the full set of assumptions of all theoretical results? [N/A] 
    \item Did you include complete proofs of all theoretical results? [N/A]
\end{itemize}
If you ran experiments...[yes]
\begin{itemize}
    \item Did you include the code, data, and instructions needed to reproduce the main experimental results (either in the supplemental material or as a URL)? [Yes] {\color{blue} Both CPU-Net's code and the training/validation data is made available in the footnote of the first page.}
     \item Did you specify all the training details (e.g., data splits, hyperparameters, how they were chosen)? [Yes] {\color{blue}The data splitting is shown in Section~\ref{sec:result}. Important model components are discussed in Section~\ref{sec:cpu_net}, and less important ones are discussed in Appendix~\ref{App:model_breakdown} along with their hyperparameters.}
     \item Did you report error bars (e.g., with respect to the random seed after running experiments multiple times)? [Yes] {\color{blue} We reported the final result with uncertainty in Section~\ref{sec:conclusion}. The uncertainty evaluation procedure is described in Appendix~\ref{App:uncertainty}. However, this model is designed to be applied to HPGe detector experiments. Therefore, learning the proper detector physics is far more important than matching two histograms.}
     \item Did you include the amount of compute and the type of resources used (e.g., type of GPUs, internal cluster, or cloud provider)? [Yes] {\color{blue} This is reported near the end of Section~\ref{subapp:hyperparam}.}
\end{itemize}
If you are using existing assets (e.g., code, data, models) or curating/releasing new assets...
\begin{itemize}
    \item If your work uses existing assets, did you cite the creators? [Yes]
    \item Did you mention the license of the assets? [N/A] {\color{blue} The dataset is self-taken at a local laboratory, we have not decided on the license of this dataset yet.}
    \item Did you include any new assets either in the supplemental material or as a URL? [Yes] {\color{blue} The dataset is included in the code repository.}
    \item Did you discuss whether and how consent was obtained from people whose data you're using/curating? [Yes]  {\color{blue} The dataset is self-taken at a local laboratory by a few authors of this work, all authors have consented to publish this result and release the data with it.}
    \item Did you discuss whether the data you are using/curating contains personally identifiable information or offensive content? [N/A]  {\color{blue} The data comes from HPGe detector and thoriated sources, no human-related or human-created content~(e.g. vocabulary, photo or video) is contained within the dataset.}
\end{itemize}
If you used crowdsourcing or conducted research with human subjects...[No] {\color{blue} All data are taken from the local HPGe detector, we did not use crowdsourcing or conducted research with human subjects.}
\begin{itemize}
    \item Did you include the full text of instructions given to participants and screenshots, if applicable? [N/A] 
    \item Did you describe any potential participant risks, with links to Institutional Review Board (IRB) approvals, if applicable? [N/A] 
    \item Did you include the estimated hourly wage paid to participants and the total amount spent on participant compensation? [N/A] 
\end{itemize}

\begin{appendix}
\section{Model Breakdown}\label{App:model_breakdown}
This appendix provides a detailed breakdown of the CPU-Net model, as requested by the NeurIPS checklist. Specifically, this section includes a detailed description of data preprocessing, model structure, hyperparameters, and training procedure. Much of the information provided in this appendix is too detailed for the scope of this work, and is therefore omitted in the main text. Details provided in this section can also be found in the CPU-Net code repository.
\subsection{Data Preprocessing}\label{App:data_preprocess}
Each raw waveform~(simulated pulses or data pulse) is first normalized into the interval between [0,1], using the following equation:
\begin{equation}
    \mathrm{WF}_{norm} = \frac{\mathrm{WF}-\min(\mathrm{WF})}{\max(\mathrm{WF})-\min(\mathrm{WF})}
\end{equation}
The normalized waveform is fed into ATN and IATN for translation purpose, where the output of each network is also normalized using the same method. In addition, we apply one-hot encoding and embedding before feeding normalized waveforms into the RNN discriminator described in Section~\ref{subapp:RNN}. The one-hot encoding is conducted through multiplying WF$_{norm}$ by 500 and rounding to the nearest integer. Then a PyTorch embedding layer is used to convert each one-hot encoding vector to an \textit{m}-dimensional embedding vector, where \textit{m} is the embedding dimension and one of the model hyperparameters.
\subsection{Positional U-Net}
\begin{figure}[htb!]
    \includegraphics[width=0.3\linewidth,trim={0pc 0pc 0pc 0pc},clip]{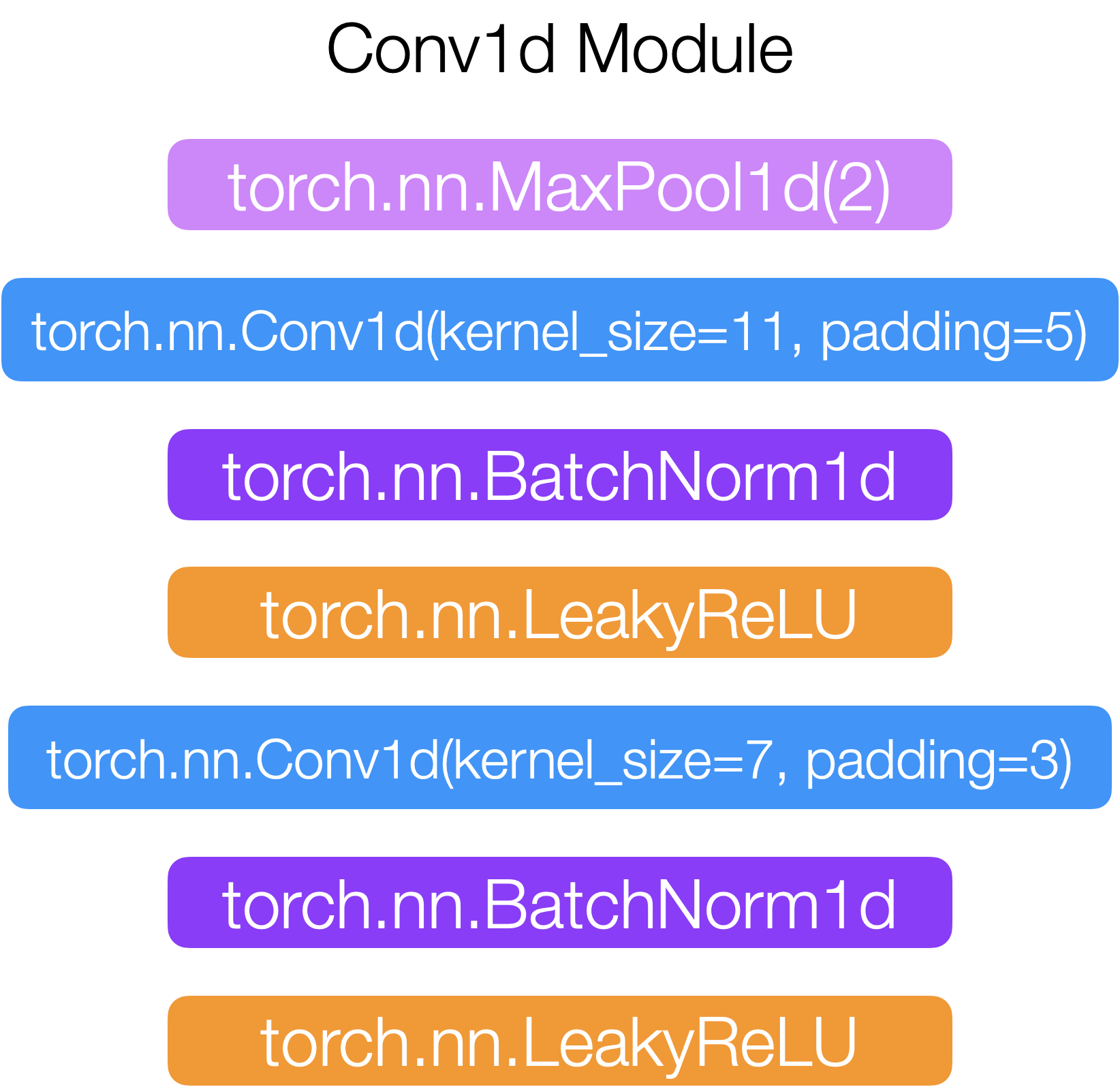}
    \hspace{0.1\linewidth}
    \includegraphics[width=0.6\linewidth,trim={0pc 0pc 0pc 0pc},clip]{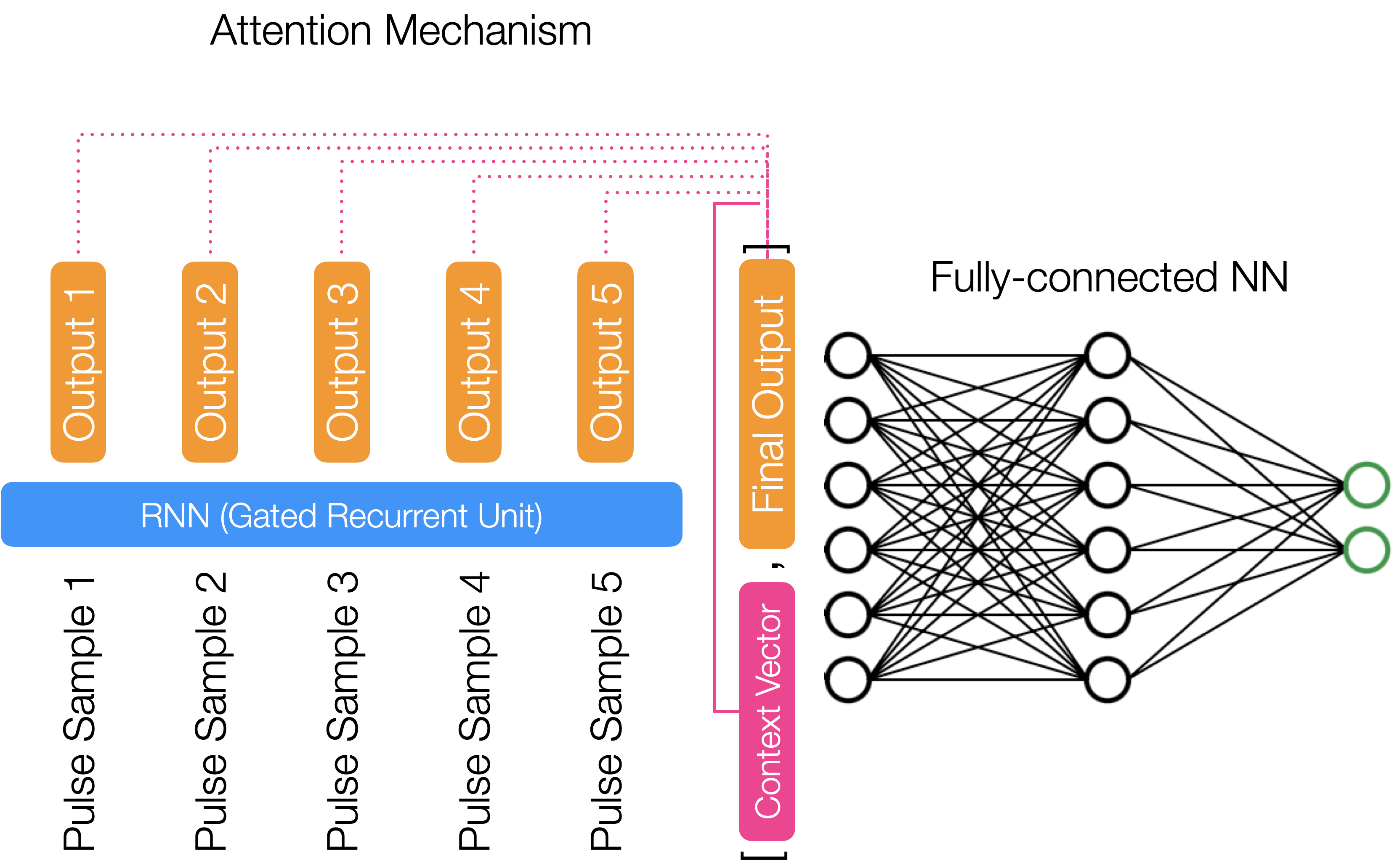}
    \caption{(Left) The layer-wise breakdown of the Conv1d Module in Figure~\ref{fig:cpunet}. (Right) The attention coupled RNN discriminator.}
    \label{fig:detail_network}
\end{figure}

The Positional U-Net structure is depicted in the left-side panel of Figure~\ref{fig:cpunet}, where the tensor shape at each stage is also denoted. The Conv1d module in Figure~\ref{fig:cpunet} is in fact a series of layers, as shown in the left-side panel of Figure~\ref{fig:detail_network}. Within this module, the kernel size is an important hyperparameter to control the reception field of CNN layers, and the padding is added to guarantee the same input and output shape. Max pooling is used in all Conv1d modules but the first one to increase the reception field. 
PU-Net's layer-wise positional encoding map $\mathcal{M}_{\mathrm{position}}$ is inspired by Reference~\cite{Transformer}. The output of 1D convolutional layer has the shape of $(\mathrm{Batch},d_{\mathrm{ch}},d_{\mathrm{seq}})$, where $d_{\mathrm{seq}}$ is the length of output feature vector and $d_{\mathrm{ch}}$ is the number of channels. $\mathcal{M}_{\mathrm{position}}$ is repeated along the batch dimension and then added to the outputs of each convolutional layer and the inputs of each upsample layer. Since those input and output tensors have different $d_{\mathrm{ch}}$ and $d_{\mathrm{seq}}$, $\mathcal{M}_{\mathrm{position}}$ has to be generated separately each time.
\begin{equation}
\mathcal{M}_{\mathrm{position}}=
\begin{pmatrix}
\sin(\frac{0}{10000^{0/d_{\mathrm{ch}}}}) &\sin(\frac{1}{10000^{0/d_{\mathrm{ch}}}})& \cdots & \sin(\frac{d_{\mathrm{seq}}}{10000^{0/d_{\mathrm{ch}}}}) \\
\cos(\frac{0}{10000^{1/d_{\mathrm{ch}}}}) &\cos(\frac{1}{10000^{1/d_{\mathrm{ch}}}})& \cdots & \cos(\frac{d_{\mathrm{seq}}}{10000^{1/d_{\mathrm{ch}}}}) \\
\sin(\frac{0}{10000^{2 /d_{\mathrm{ch}}}}) &\sin(\frac{1}{10000^{2/d_{\mathrm{ch}}}})& \cdots & \sin(\frac{d_{\mathrm{seq}}}{10000^{2/d_{\mathrm{ch}}}}) \\
\vdots&\vdots&\ddots&\vdots\\
\cos(\frac{0}{10000^{d_{\mathrm{ch}}/d_{\mathrm{ch}}}}) &\cos(\frac{1}{10000^{d_{\mathrm{ch}}/d_{\mathrm{ch}}}})& \cdots & \cos(\frac{d_{\mathrm{seq}}}{10000^{d_{\mathrm{ch}}/d_{\mathrm{ch}}}}) \\
\end{pmatrix}
\label{eqn:pos_encoding}
\end{equation}

Lastly, a reparameterization trick is added to the bottom of PU-Net. The trick was proposed by the Variational Autoencoder paper~\cite{VAE} to sample from random space while preserving the gradient flow. Reference~\cite{AAE} pointed out that this trick has the ability to increase the stochasticity of machine learning models. Based on our experimental outcomes, an increased stochasticity helps the ATN and IATN to learn the reconstruction parameters' distributions better. Therefore, we decided to add this method to the latent space vector of PU-Net. Unlike Reference~\cite{VAE}, we did not use  Kullback-Leibler Divergence to regulate the reparameterized random distribution.
\subsection{RNN Discriminator}~\label{subapp:RNN}
The RNN discriminator is depicted in the right-side panel of Figure~\ref{fig:detail_network}. A single-layer bidirectional RNN is used as the discriminator model in CPU-Net. We adopt the Gated Recurrent Unit~\cite{GRU} as the internal structure of the RNN. The input raw pulses are first embedded with $m=128$, and the RNN yields a 64-dimensional output $\Vec{I}(t)$ at each intermediate step $t$ as well as a final output  $\Vec{F}$. We then use an attention mechanism~\cite{attention} to boost the performance of RNN discriminator. The attention mechanism contains an attention matrix A of dimension (64,64), which is used to calculate the attention score between $\Vec{F}$ and each $\Vec{I}(t)$:
\begin{equation}
    s(t) = \mathrm{Softmax}[\Vec{I}(t) A \Vec{F}]
\end{equation}
A context vector is produced by sum $\Vec{I}(t)$ with the weight $s(t)$ at each $t$. Finally, the context vector and the final output vector is concatenated and fed into a series of fully connected layers to produce a single output.
\subsection{Training and Hyperparameter Search}\label{subapp:hyperparam}
Three losses are optimized simultaneously for the simulated pulse translation path in Figure~\ref{fig:cpunet}:
\begin{align}
    &L_{\mathrm{Cycle}} = \lvert X - \Bar{\Lambda}(\Lambda(X))\lvert\label{cycle}\\
    &L_{\mathrm{Adversarial}} = E_{\mathcal{X'}}\log(\delta(X')) - E_{\Lambda(\mathcal{X})}\log(1 - \delta(\Lambda(X)))\label{adv}\\
    &L_{\mathrm{Identity}} = \lvert X' - \Lambda(X')\lvert\label{adv}
\end{align}
Besides the cycle loss and adversarial discussed in Section~\ref{sec:cpu_net}, there is an additional loss $L_{\mathrm{Identity}}$ that was not discussed. In the CPU-Net structure, the ATN $\Lambda$ is obligated to produce data-pulse-like events at the output end. $L_{\mathrm{Identity}}$ ensures that the ATN preserves its shape when an actual detector pulse $X'$ is fed into $\Lambda$. An additional three complimentary losses are defined for the detector pulse translation path. Therefore, a total of six losses are optimized simultaneously.

ADAM~\cite{ADAM} optimizers are used for all losses. The total training cycle contains 3000 batches with batch size 16. The learning rate decays linearly from 0.001 to 0 with a step size of 500. A single training trial takes about 0.5 GPU hours on NVIDIA A100 GPU. Hand-tuning of hyperparameters is performed to obtain the most stable and efficient version of CPU-Net.
\section{Uncertainty Evaluation}\label{App:uncertainty}
This section describes the uncertainty evaluation procedure of this work. We used the Intersection over Union~(IoU) measurement between the data histogram and ATN output histogram to quantify the performance of translation tasks. HIoU is calculated by iterating through each bin of both histograms, summing up the bin-wise minimum as the intersection and the bin-wise maximum as the union. HIoU varies between 0-100\%, where 0\% indicates no intersection and 100\% indicates complete overlap between the two histograms.

\begin{figure}[htb!]
    \includegraphics[width=0.45\linewidth,trim={3pc 3pc 7pc 7pc},clip]{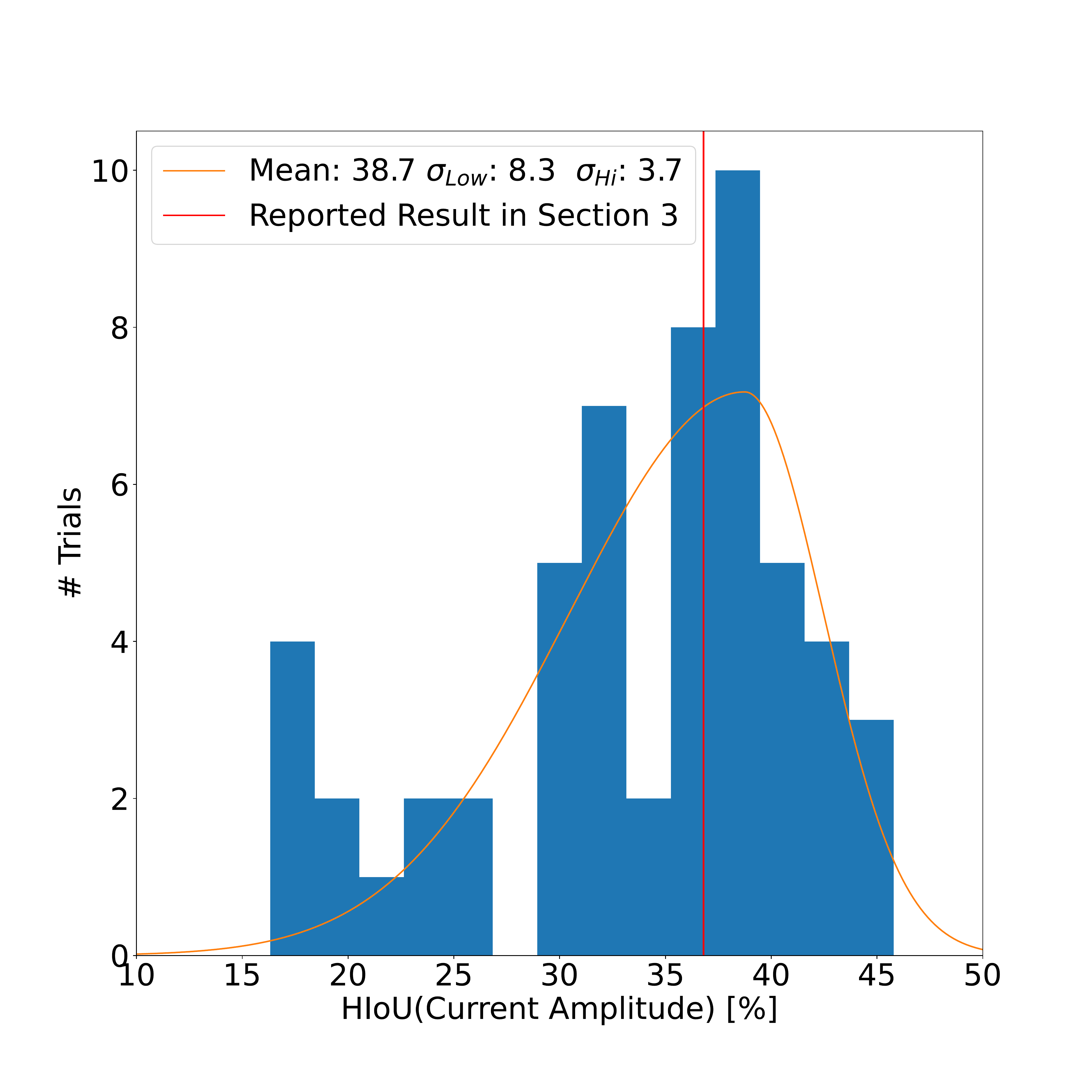}
    \hspace{0.05\linewidth}
    \includegraphics[width=0.45\linewidth,trim={3pc 3pc 7pc 7pc},clip]{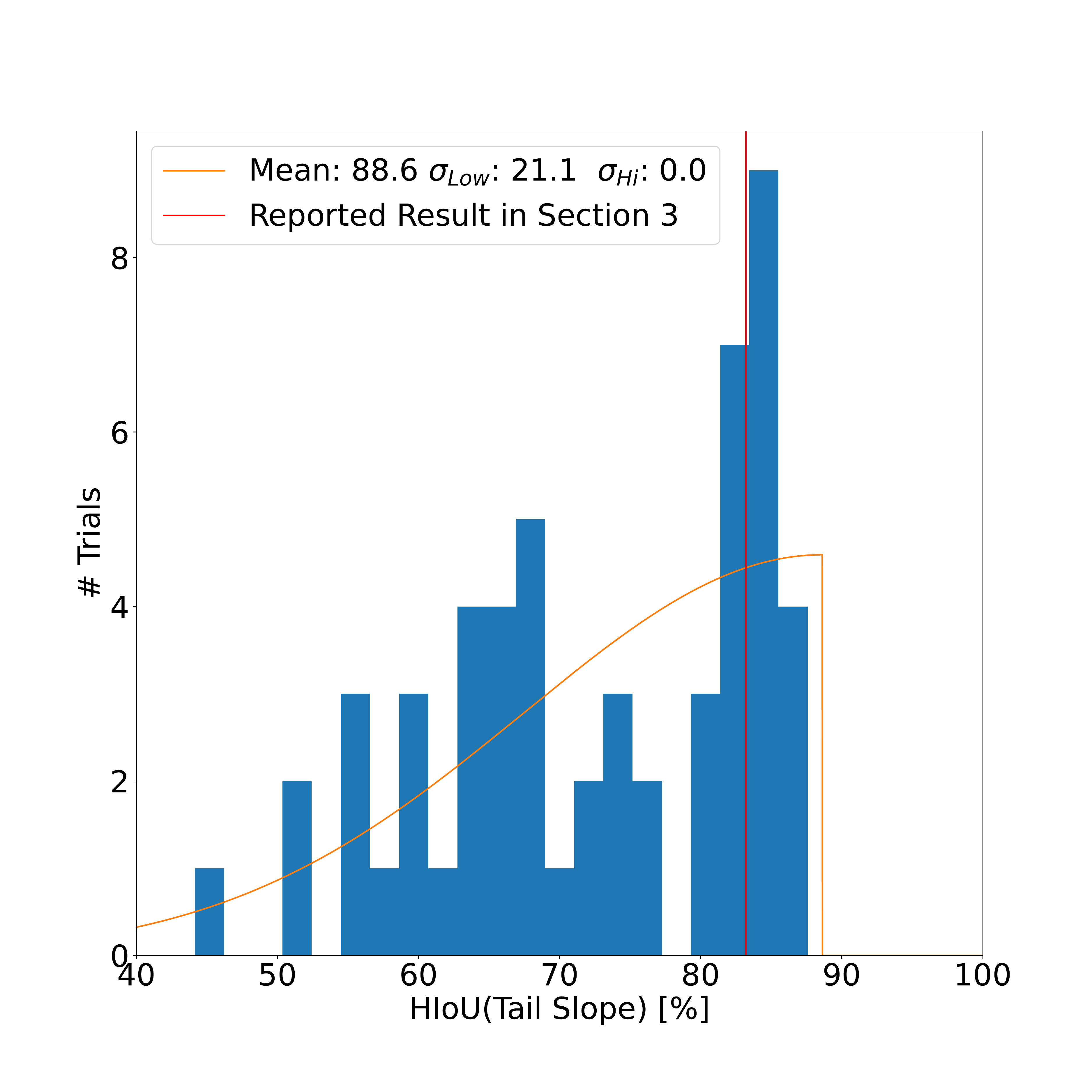}
    \caption{Histogram Intersection over Union~(HIoU) distribution of CPU-Net trials over maximal current amplitude $I_{max}$~(Left) and tail slope $c_{tail}$~(Right). }
    \label{fig:uncert}
\end{figure}

To calculate uncertainty, we first trained 56 trials of CPU-Net using different random seeds. The HIoU over $I_{max}$ and $c_{tail}$ are then calculated for each trial over the same validation dataset of 3000 pulses. We then fit Fechner's Distribution to obtain the mean and the lower- and upper-statistical uncertainty. As shown in Figure~\ref{fig:uncert}, some trials have much lower HIoU than the mean value due to failed GAN training in CPU-Net, as discussed in Section~\ref{sec:broader_impact} Limitations. Therefore the lower-statistical uncertainty is always larger than the upper-statistical uncertainty. We then quote the difference between the reported result~(red vertical line) and the fitted mean as the systematic uncertainty. Since the deviation from distribution mean could come from both statistical and systematic effects, treating it entriely as systematic uncertainty is a conservative estimation. Finally, we add statistical and systematic uncertainties in quadrature to report the final uncertainty in Section~\ref{sec:result}.
We want to point out that HIoU is not the most proper way to evaluate CPU-Net's performance. Since CPU-Net will be applied to HPGe detector experiments, learning the proper detector physics is far more important than matching two histograms. As we showed in Section~\ref{sec:result} and Figure~\ref{fig:result} Top, CPU-Net has learned the missing detector physics in simulation~(i.e. 0 preamplifier integration time in \texttt{siggen}) and applies proper corrections by learning from data. This, in turn, delivers an increased HIoU on critical reconstruction parameters. 
\end{appendix}
\end{document}